\documentclass[prd,showpacs,preprint]{revtex4}
\usepackage{graphicx,color,epsfig}

\newcommand{\ba}{\begin{array}}
\newcommand{\ea}{\end{array}}
\newcommand{\bd}{\begin{displaymath}}
\newcommand{\ed}{\end{displaymath}}
\newcommand{\be}{\begin{equation}}
\newcommand{\ee}{\end{equation}}
\newcommand{\beq}{\begin{eqnarray}}
\newcommand{\eeq}{\end{eqnarray}}

\newcommand{\non}{\nonumber\\ }
\newcommand{\psl}{ p \hspace{-1.8truemm}/ }
\newcommand{\nsl}{ n \hspace{-2.2truemm}/ }
\newcommand{\vsl}{ v \hspace{-2.2truemm}/ }

\begin{document}

\title{The study of  $B\to J/\Psi~\eta^{(\prime)}$ decays and
determination\\ of $\eta-\eta^{\prime}$ mixing angle}

\author{Jing-Wu Li$^{1}${\footnote {Email:lijw@email.xznu.edu.cn}},~~~Dong-Sheng Du$^{2}${\footnote {Email:duds@mail.ihep.ac.cn}}}

\vspace*{1.0cm}

\affiliation{$^{1}$Department of Physics, Xu Zhou Normal University,
XuZhou 221116, China,\\$^{2}$Institute of High Energy Physics, P.O.
Box 918(4), Beijing 100049, China}


\vspace*{1.0cm}

\date{\today}
\begin{abstract}
We study $B\to J/\Psi~\eta^{(\prime)}$ decays and suggest two
methods to determine the $\eta-\eta^{\prime}$ mixing angle. We
calculate not only the factorizable contribution in QCD facorization
scheme but also  the nonfactorizable hard spectator corrections in
pQCD approach. We  get the branching ratio of $B\to J/\Psi~\eta$
which is consistent with recent experimental data and predict the
branching ratio of $B\to J/\Psi~\eta^{\prime}$ to be $7.59\times
10^{-6}$. Two methods for determining $\eta-\eta^{\prime}$ mixing
angle are suggested in this paper. For the first method, we get the
$\eta-\eta^{\prime}$ mixing angle  to be about $-13.1^{\circ}$,
which is in consistency with others in the literature. The second
 method depends on less parameters so can be used to determine
 the $\eta-\eta^{\prime}$ mixing angle with better accuracy but
 needs, as an input, the branching ratio for $B\to
 J/\Psi~\eta^{\prime}$ which should be measured in the near future.

\end{abstract}

\pacs{13.25.Hw, 12.38.Bx} \maketitle


\section{Introduction}

Nonleptonic decays of B mesons is a good place for testing the
Standard Model and finding new physics beyond the SM. Several useful
methods have been created to calculate the hadronic nonleptonic
decay matrix elements, such as the naive factorization
assumption(NF)\cite{BSW,fac}, the pQCD approach\cite{PQCD}, the
QCD-improved factorization (QCDF)\cite{BBNS}, soft collinear
effective theoty (SCET)\cite{SCET} etc. Most of the predictions from
these methods are consistent with experimental data, but  these
methods do not apply to B mesons decays into charmonia\cite{a1a2},
such as $B\rightarrow J/\Psi K$. The branching ratio of
$B\rightarrow J/\Psi K$ from NF is too small compared to the
experimental data by Babar \cite{Babar-hep04},
 \begin{eqnarray}
Br(B^+ \to J/\psi K^+) &=& (10.61\pm 0.15 \pm 0.48) \times 10^{-4}
\, ,\nonumber \\
Br(B^0 \to J/\psi K^0) &=& (8.69 \pm 0.22 \pm 0.30) \times 10^{-4}
\, , \label{BRexpjpsaik}
\end{eqnarray}
   The large inconsistency between prediction from Naive Factorization  and experimental data shows
    that nonfactorizable contribution may play an important role . Some other approaches  have been tried
to solve the puzzle\cite{LRev}. The prediction  $B(B^0 \to J/\psi
K^0)\approx 1\times 10^{-4}$ from QCD-improved factorization(QCDF)
is too small to account for  the data. In the calculation  of  the
hard spectator scattering diagrams by QCDF, logarithmical
divergences are generated from the end-point region. So to make an
estimation, arbitrary cutoffs for parameterizing  the divergence
have been introduced, which render the contribution of the
nonfactorizable hard spectator scattering diagrams out of control. A
method to calculate the hard spectator scattering diagrams was
introduced by the authors of Ref.\cite{cljpsi}. This method can give
good explanation for the decays of $B\rightarrow J/\Psi K$.

  The  $B\to J/\Psi~\eta^{(\prime)}$ decays were calculated with pQCD
  approach in Ref.~\cite{xiaoeta}. The predicted central value of $B\to J/\Psi~\eta$
   in Ref.~\cite{xiaoeta} is four times smaller than the  recent measured
   one by Belle\cite{bellejpsi}. The reason is that the characteristic scale in
   the factorizable diagram of $B\to J/\Psi~\eta^{(\prime)}$ is around
   1 GeV, which means that the pQCD approach can not apply. The calculation of many B decays
   into charmonia shows that the method in \cite{cljpsi} is applicable for calculating the
decay amplitude of
    $B\to J/\Psi~\eta^{(\prime)}$. In this paper, we  calculate
    the $B\to J/\Psi~\eta^{(\prime)}$ decays
    with the methods put forward in \cite{cljpsi}.

The mixing of $\eta $ and $\eta\prime$ and their components are
interesting topics to be investigated. Many attempts have been made
to determine the mixing angle and the gluonic component
\cite{ekou01}-\cite{huangcao}. Most of the  authors obtained the
mixing angle $\theta_p$ in the range between $-20^{\circ}$ to
$-10^{\circ}$ by fitting the experimental data . The
$\eta-\eta^{\prime}$ mixing angle is generally determined
 through the calculation of the decay amplitudes or
transition form factor, so the determination of $\eta-\eta^{\prime}$
mixing angle depends on the choice of some uncertain parameters and
assumption about the variation of the form factor with momentum
transfer $Q^2$, the decay constants of $\eta$ and $\eta^{\prime}$,
CKM matrix elements, and the choice of model-dependent wave function
of the relevant mesons . That  means that the fitted mixing angle
has many uncertainty sources,  such as the chiral enhancement scale
$m_0^{\eta^{\bar{d}d}}$ chosen from 1.07 GeV to 1.5 GeV , the
uncertain shape parameter in the wave function which causing big
uncertainties of the decay rates and the model-dependent wave
functions of the relevant mesons . We think that it is not a good
way to determine $\eta-\eta^{\prime}$ mixing angle with too many
parameters and assumptions. So we  try to find a better method to
determine it. Based on only one assumption that the decay
 constant and the distribution
amplitude  of the $d\bar{d}$ component for $\eta$
  is the same as that for  $\eta^{\prime}$\cite{feldmann1,feldmann,ball},
  we can derive the relation between
  the branching ratios of $B\to J/\Psi~\eta^{(\prime)}$.
From this relation, we can determine $\eta-\eta^{\prime}$ mixing
angle. The only inputs we need are the masses of $\eta^{(\prime)}$
and the experimental value of the ratio of the branching ratios for
$B\to J/\Psi~\eta^{(\prime)}$. The masses of $\eta^{(\prime)}$ have
small uncertainties\cite{pdg2004}, so the  $\eta-\eta^{\prime}$
mixing angle determined in this way has much less uncertainty
sources.

 This paper is organized as follows. In Sec.~II, we derive the
formulas for the amplitudes of the $B^0 \to J/\Psi~\eta^{(\prime)}$.
Two methods for determining the $\eta-\eta^{\prime}$ mixing angle
are presented. Summary is given in Sec.~III . Some input parameters
and mesons wave function are listed in the appendix in Sec.~IV.

\section{Branching Ratios for the Decays of $B^0 \to J/\Psi~\eta^{(\prime)}$}
\label{br}

The accuracy of the mixing angle of $\eta$ and $\eta^{\prime}$
depend on the reliability of the method for calculating the decay
amplitudes. $B\to J/\Psi \eta^{(\prime)}$ are good processes for
determining the $\eta-\eta^{\prime}$ mixing angle. From the
prediction for $B\to J/\Psi K^{(\ast)}$ in Ref.~\cite{cljpsi} and
that for $B\to J/\Psi \pi^{0}$ in our paper\cite{jpsipiz}, we
believe that the method in Ref.~\cite{cljpsi} can be used to
calculate the branching ratios of $B^0\to J/\Psi \eta^{(\prime)}$ so
as to determine the $\eta-\eta^{\prime} $ mixing angle.

The $\eta$ and $\eta^\prime$  are neutral pseudoscalar ($J^P=0^-$)
mesons. There are two different mixing scheme to describe the $\eta$
-$\eta^\prime$ mixing, we choose the mixtures of the $SU(3)_F$
singlet $\eta_1$ and the octet $\eta_8$\cite{feldmann1,feldmann}:
\beq \left(\begin{array}{c}
     \eta \\ \eta^{\prime} \end{array} \right)
= \left(\begin{array}{cc}
 \cos{\theta_p} & -\sin{\theta_p} \\
 \sin{\theta_p} & \cos{\theta_p} \\ \end{array} \right)
 \left(\begin{array}{c}
 \eta_8 \\ \eta_1 \end{array} \right),
\label{eq:e-ep} \eeq with \beq \eta_8&=&\frac{1}{\sqrt{6}}\left (
u\bar{u}+d\bar{d}-2s\bar{s}\right ),\non
\eta_1&=&\frac{1}{\sqrt{3}}\left (u\bar{u}+d\bar{d}+s\bar{s}\right
), \label{eq:e1-e8} \eeq where $\theta_p$ is the
$\eta-\eta^{\prime}$ mixing angle .

In order to determine the mixing angle of $\eta$ and $\eta\prime$,
we choose to calculate the $B^0 \to J/\Psi~\eta^{(\prime)}$ decays.

For the $B^0 \to J/\Psi~\eta^{(\prime)}$ decays, the effective
Hamiltonian is given by \cite{buras96},
\begin{eqnarray}
H_{\rm eff}&=&\frac{G_F}{\sqrt
2}\left\{V_{cb}V_{cd}^{*}[C_1(\mu)O_1+
C_2(\mu)O_2]-V_{tb}V_{td}^{*}\sum_{k=3}^{10}C_k(\mu)O_k\right\}\;,
\label{effh}
\end{eqnarray}
with the Cabibbo-Kobayashi-Maskawa (CKM) matrix elements $V$ and the
four-fermion operators,
\begin{eqnarray}
& &O_1 = (\bar{d}_ic_j)_{V-A}(\bar{c}_jb_i)_{V-A}\;,\;\;\;\;\;\;\;\;
O_2 = (\bar{d}_ic_i)_{V-A}(\bar{c}_jb_j)_{V-A}\;,
\nonumber \\
& &O_3 =(\bar{d}_ib_i)_{V-A}\sum_{q}(\bar{q}_jq_j)_{V-A}\;,\;\;\;\;
O_4 =(\bar{d}_ib_j)_{V-A}\sum_{q}(\bar{q}_jq_i)_{V-A}\;,
\nonumber \\
& &O_5
=(\bar{d}_ib_i)_{V-A}\sum_{q}(\bar{q}_jq_j)_{V+A}\;,\;\;\;\;O_6
=(\bar{d}_ib_j)_{V-A}\sum_{q}(\bar{q}_jq_i)_{V+A}\;,
\nonumber \\
& &O_7
=\frac{3}{2}(\bar{d}_ib_i)_{V-A}\sum_{q}e_q(\bar{q}_jq_j)_{V+A}\;,
\;\; O_8
=\frac{3}{2}(\bar{d}_ib_j)_{V-A}\sum_{q}e_q(\bar{q}_jq_i)_{V+A}\;,
\nonumber \\
& &O_9
=\frac{3}{2}(\bar{d}_ib_i)_{V-A}\sum_{q}e_q(\bar{q}_jq_j)_{V-A}\;,
\;\; O_{10}
=\frac{3}{2}(\bar{d}_ib_j)_{V-A}\sum_{q}e_q(\bar{q}_jq_i)_{V-A}\;,
\end{eqnarray}
$i, \ j$ being the color indices.

In this paper, we take the light-cone coordinates $(p^+, p^-, {\bf
p}_T)$ to describe the four-dimensional momenta of the meson,
\begin{eqnarray} p^\pm = \frac{1}{\sqrt{2}} (p^0 \pm p^3), \quad and \quad
{\bf p}_T = (p^1, p^2).
 \end{eqnarray}

At the rest frame of the B meson, the momentum $P_{1}$ of the B
meson is
\begin{eqnarray}
  P_1 &=& \frac{M_B}{\sqrt{2}} (1,1,{\bf 0}_T)
\end{eqnarray}

the $J/\Psi (\eta)$ meson momentum $P_{2}(P_{3})$ can be written as

\begin{eqnarray}
  \quad P_2 =
\frac{M_B}{\sqrt{2}}(1-r_3^2,r_2^2,{\bf 0}_T), \quad P_3 =
\frac{M_B}{\sqrt{2}} (r_3^2,1-r_2^2,{\bf 0}_T)
\end{eqnarray}

with $r_2=m_{J/\psi}/m_B$, $r_3=m_{\eta}/m_B$.

The polarization  vectors of the $J/\Psi$ meson are parameterized as
\begin{eqnarray}
\epsilon_{2L}=\frac{1}{\sqrt{2}r_2}\left(1,-r_2^2, {\bf
0}_T\right)\;,\;\;\;\; \epsilon_{2T}=\left(0,0, {\bf
1}_T\right)\;.\label{pol}
\end{eqnarray}

The decay width of of $B^0 \to J/\Psi~\eta$ is

\begin{equation}\label{gamma}
\Gamma=\frac{1}{32\pi m_B}G_F^2(1-r_2^2+\frac{1}{2}r_{2}^4-r_3^2)
|{\cal A}|^2\;.
\end{equation}
  The amplitude ${\cal A}$ consists of  factorizable part and
  nonfactorizable part.  It can be written as
\begin{eqnarray}
{\cal A}&=&f^{\eta}_{d\bar{d}}(A_{FA}+ A_{VERT}+ A_{HS})\;,
\label{ampeta}
\end{eqnarray}

with mixing factor
\begin{equation}\label{fd}
    f^{\eta}_{d\bar{d}}=\frac{1}{\sqrt{6}}\cos{\theta_P}-\frac{1}{\sqrt{3}}\sin{\theta_P}
\end{equation}

where $A_{FA}$ denotes the factorizable contribution, $A_{VERT}$ is
the vertex corrections from Fig.~\ref{nonfc}.(a)-(d), $A_{HS}$ is
the hard spectator scattering correction from
Fig.~\ref{nonfc}.(e)-(f).

\subsection{Factorizable Contribution and Vertex Correction In QCDF}
\begin{figure}[tb]
\begin{center}
\epsfig{file=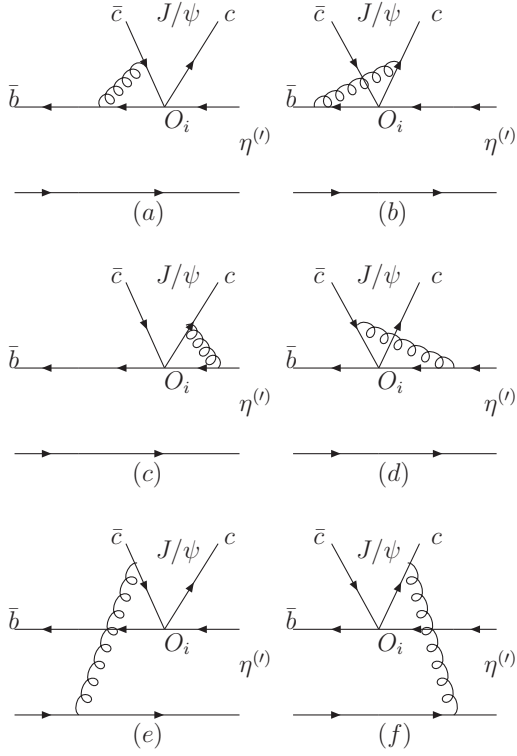,height=10cm}
\caption{ Nonfactorizable contribution to the $B^0\to
J/\Psi~\eta^{(\prime)}$ decays} \label{nonfc}
\end{center}
\end{figure}

The factorizable part $A_{FA}$ of amplitude ${\cal A}$ in
Eq.~(\ref{ampeta}) for $B\to J/\Psi \eta$ decay  can not be
 calculated reliably in pQCD approach, because its characteristic scale is around 1 GeV \cite{cljpsi}.
 We here compute the factorizable part of amplitude and the vertex correction
from Fig.~\ref{nonfc}.(a)-(d) in QCDF \cite{BBNS} instead of pQCD
approach and get

\begin{eqnarray}
A_{FA}+A_{VERT}=a_{eff} m_B^2 f_{J/{\psi}}F_1^{B\to
\eta}(m_{J/\psi}^2) (1-r_2^2)\;,\label{amppinf}
\end{eqnarray}
where $f_{J/{\psi}}$ is decay constant of $J/\psi$ meson, $F_1^{B\to
\eta}$ is the $B \to \eta$ transition form factor defined as
\begin{eqnarray}
\langle \eta(P_3)|{\bar b}\gamma_\mu d|B(P_1)\rangle =F_1^{B\to
\eta}(q^2)\left[(P_1+P_3)_\mu-\frac{m_B^2-m_\eta^2}{q^2}q_\mu\right]
+F_0^{B\to \eta}(q^2)\frac{m_B^2-m_\eta^2}{q^2}q_\mu\;,\label{fp}
\end{eqnarray}
$q=P_1-P_3$ being the momentum transfer, and $m_\eta$ the $\eta$
meson mass.

The Wilson coefficient $a_{eff}$ for $B^0\to J/\Psi \eta$ can be
derived in QCDF\cite{qcdfvc},
\begin{eqnarray}
a_{eff}&=&V_c^\ast \left[C_1+V_c^\ast
\frac{C_2}{N_c}+\frac{\alpha_s}{4\pi}\frac{C_F}{N_c}C_2
\left(-18+12\ln\frac{m_b}{\mu}+f_I\right)\right]\nonumber\\&&-V_t^\ast
\Big[C_3+\frac{C_4}{N_c}+\frac{\alpha_s}{4\pi}\frac{C_F}{N_c}C_4
\left(-18+12\ln\frac{m_b}{\mu}+f_I\right)\nonumber\\&&+C_5+\frac{C_6}{N_c}+\frac{\alpha_s}{4\pi}\frac{C_F}{N_c}C_6
\left(6-12\ln\frac{m_b}{\mu}-f_I\right)+C_7+\frac{C_8}{N_c}+C_9+\frac{C_{10}}{N_c}\Big]\;,\nonumber
\label{anfandver}
\end{eqnarray}

with the function,
\begin{eqnarray}
f_I=\frac{2\sqrt{2N_c}}{f_{J/\psi}}\int
dx_3\Psi^L(x_2)\left[\frac{3(1-2x_2)}{1-x_2}\ln x_2-3\pi
i+3\ln(1-r_2^2)+\frac{2r_2^2(1-x_2)}{1-r_2^2 x_2}\right]\;,
\end{eqnarray}

For the $B\to \eta$ transition form factor, we employ the models
derived from the light-cone sum rules \cite{ballform}, which is
parameterized as
\begin{eqnarray}
F_1^{B\to
\eta}(q^2)=\frac{r_1}{1-q^2/m_1}+\frac{r_2}{(1-q^2/m_1^2)^2}\;\label{f1eta}
\end{eqnarray}

with $r_1=0.122$, $r_2=0.155$, $m_1=5.32Gev$, for $B \to \eta$
transition.

\subsection{Hard Spectator Scattering Corrections In pQCD Approach}
For the  hard spectator scattering corrections $A_{HS}$ from
Fig.~\ref{nonfc}.(e)-(f), QCDF is not appropriate due to the
end-point singularity from vanishing parton momenta. The
nonfactorizable contribution has a characteristic hard scale higher
than that in B meson transition form factor~\cite{Chou:2001bn}.
Therefore, we can employ pQCD approach based on $k_T$ factorization
theorem, which is free of the end-point singularity for the
spectator amplitude \cite{cljpsi}. The nonfactorizable hard
spectator amplitudes can be written as,
\begin{eqnarray}
A_{HS}&=&V_c^\ast {\cal M}_{1}^{(J/\psi\eta)}- V_t^\ast {\cal
M}_{4}^{(J/\psi \eta)}-V_t^\ast{\cal M}_{6}^{(J/\psi
\eta)}\;,\label{amppihs}
\end{eqnarray}
where the amplitudes ${\cal M}_{1,4}^{(J/\psi \eta)}$ and ${\cal
M}_{6}^{(J/\psi \eta)}$ result from the $(V-A)(V-A)$ and
$(V-A)(V+A)$ operators in Eq.~(\ref{effh}), respectively. Their
factorization formulas are given by pQCD approach. In the
calculation of ${\cal M}_{1,4}^{(J/\psi \eta)}$ and ${\cal
M}_{6}^{(J/\psi \eta)}$, because $J/\psi$ is heavy,  we reserve  the
power terms of $r_2$ up  to $\mathcal{O}(r^{4}_{2})$, the power
terms of $r_3$ up to $\mathcal{O}(r^{2}_{3})$ . When  $r_2^{4}$ and
$r_3^{2}$ are taken as zero, the  ${\cal M}_{1,4}^{(J/\psi
\eta)}/m_{B}^{2}$ , ${\cal M}_{6}^{(J/\psi \eta)}/m_{B}^{2}$  in
this paper can be  reduced to the corresponding ${\cal
M}_{1,4}^{(J/\psi \eta)}$ , ${\cal M}_{6}^{(J/\psi \eta)}$  in
ref.\cite{cljpsi},
\begin{eqnarray}
{\cal M}_{1,4}^{(J/\psi \eta)} &=&16\pi m_B^2 C_{F}\sqrt{2N_{c}}%
\int_{0}^{1}[dx]\int_{0}^{\infty }b_{1}db_{1} \Phi _{B}(x_{1},b_{1})
\nonumber \\
&& \times \Big\{ \Big[ (1-2r^{2}_{2}+r^{4}_{2})(1-x_{2}) \Phi _{\pi}
( x_{3} )\Psi^{L}(x_{2})
+\frac{1}{2} (r^{2}_{2}-r^{4}_{2}) \Phi_{\eta}(x_{3})\Psi^{t }(x_{2}) \nonumber \\
&&- r_{\eta} (1-r^2_2)x_3 \Phi_{\eta}^P(x_3) \Psi _{L}(x_{2})+
r_{\eta} \left( 2r^{2}_{2}(1-x_2)+(1-r^2_2)x_3 \right)
\Phi_{\eta}^t(x_3) \Psi^{L}(x_{2})
  \Big]\nonumber \\
&&\times E_{1,4}(t_d^{(1)})h_d^{(1)}(x_1,x_2,x_3,b_1)
\nonumber \\%
&& - \Big[ (x_2-x_2r^{4}_{2}+x_3-2r^2_{2}x_3+r^{4}_{2}x_3)\Phi
_{\pi} ( x_{3}
)\Psi^{L}(x_{2})+r^{2}_{2}(2r_{\eta}\Phi_{\eta}^t(x_{3})\nonumber
\\ &&-\frac{1}{2}(1-r^{2}_{2})\Phi_{\eta}(x_{3}))\Psi^{t}(x_{2})
-r_{\eta} (1-r^2_2)x_3 \Phi_{\eta}^P(x_3) \Psi _{L}(x_{2})\nonumber
\\ &&- r_{\eta}
\left( 2r^{2}_{2}x_2+(1-r^2_2)x_3 \right) \Phi_{\eta}^t(x_3)
\Psi^{L}(x_{2})\Big]\nonumber \\
&& \times E_{1,4}(t^{(2)}_d)
h_d^{(2)}(x_1,x_2,x_3,b_1)\;,\label{psi4}\\
{\cal M}_{6}^{(J/\psi \eta)} &=&16\pi m_B^2 C_{F}\sqrt{2N_{c}}%
\int_{0}^{1}[dx]\int_{0}^{\infty }b_{1}db_{1} \Phi _{B}(x_{1},b_{1})
\nonumber \\
&& \times \Big\{  \Big[
(1-x_2+r^4_{2}x_2+x_3-2r^2_{2}x_3+r^4_{2}x_3-r^4_{2})\Phi _{\pi} (
x_{3} )\Psi^{L}(x_{2})\nonumber \\
&& +
r^{2}_{2}(2r_{\eta}\Phi_{\eta}^t(x_{3})-\frac{1}{2}(1-r^{2}_{2})\Phi_{\eta}(x_{3}))\Psi^{t}(x_{2})
 \nonumber \\
&& -r_{\eta} (1-r^2_2)x_3 \Phi_{\eta}^P(x_3) \Psi^{L}(x_{2})-
r_{\eta} \left( 2r^{2}_{2}(1-x_2)+(1-r^2_2)x_3 \right)
\Phi_{\eta}^t(x_3)
\Psi^{L}(x_{2})\Big]\nonumber \\
&&\times E_{6}(t_d^{(1)})h_d^{(1)}(x_1,x_2,x_3,b_1)
\nonumber \\%
&& - \Big[ (1-2r^{2}_{2}+r^{4}_{2})x_{2} \Phi _{\pi} ( x_{3}
)\Psi^{L}(x_{2})
+ \frac{1}{2} (r^{2}_{2}-r^{4}_{2}) \Phi_{\eta}(x_{3})\Psi^{t }(x_{2}) \nonumber \\
&&- r_{\eta} (1-r^2_2)x_3 \Phi_{\eta}^P(x_3) \Psi^{L}(x_{2})+
r_{\eta} \left( 2r^{2}_{2}x_2+(1-r^2_2)x_3 \right)
\Phi_{\eta}^t(x_3) \Psi^{L}(x_{2})
  \Big]\nonumber \\
&& \times E_{6}(t^{(2)}_d) h_d^{(2)}(x_1,x_2,x_3,b_1)
\Big\}\;,\label{psi6}
\end{eqnarray}
with the color factor $C_F=4/3$, the number of colors $N_c=3$, the
symbol $[dx]\equiv dx_1 dx_2 dx_3$ and the mass ratio
$r_{\eta}=m_0^{\eta^{\bar{d}d}}/m_B$, $m_0^{\eta^{\bar{d}d}}$ being
the chiral scale associated with the $\eta$ meson.

In the derivation of spectator correction in pQCD approach, we need
to input the wave function of relevant mesons  , we list the wave
functions in appendix.

The evolution factors are written as\cite{cljpsi}
\begin{eqnarray}
E_{i}(t) &=&\alpha _{s}(t) a_{i}^{\prime}(t)S(t)|_{b_{3}=b_{1}}\;,
\end{eqnarray}
with the Wilson coefficients,
\begin{eqnarray}
a_{1}^{\prime } &=&\frac{C_{2}}{N_{c}};, \nonumber\\
a_{4}^{\prime } &=&\frac{1}{N_{c}} \left(
C_{4}+\frac{3}{2}e_{c}C_{10}\right)\;, \nonumber\\
a_{6}^{\prime} &=&\frac{1}{N_{c}}\left(
C_{6}+\frac{3}{2}e_{c}C_{8}\right)\;.
\end{eqnarray}
The Sudakov exponent is given by\cite{cljpsi}
\begin{eqnarray}
S(t)&=&S_B(t)+S_K(t)\;,\nonumber\\
S_{B}(t)&=&\exp\left[-s(x_{1}P_{1}^{+},b_{1})
-\frac{5}{3}\int_{1/b_{1}}^{t}\frac{d{\bar{\mu}}} {\bar{\mu}} \gamma
(\alpha _{s}({\bar{\mu}}))\right]\;,
\label{sb} \\
S_{K}(t)&=&\exp\left[-s(x_{3}P_{3}^{-},b_{3})
-s((1-x_{3})P_{3}^{-},b_{3})
-2\int_{1/b_{3}}^{t}\frac{d{\bar{\mu}}}{\bar{\mu}} \gamma
(\alpha_{s}({\bar{\mu}}))\right]\;, \label{sbk}
\end{eqnarray}

The hard functions $h_d^{(j)}$, $j=1$ and 2, are
\begin{eqnarray}
h^{(j)}_d&=& \frac{1}{D-D_j} \left( \begin{array}{cc}
 K_{0}(\sqrt{D_{j}}m_Bb_{1})-K_{0}(\sqrt{D}m_Bb_{1}) &  \mbox{for $D_{j} \geq 0$}  \\
 \frac{i\pi}{2} H_{0}^{(1)}\left(\sqrt{|D_{j}|}m_Bb_{1}\right)-K_{0}(\sqrt{D}m_Bb_{1})
   & \mbox{for $D_{j} < 0$}
  \end{array} \right)\;,
\label{hjd}
\end{eqnarray}
with the arguments,
\begin{eqnarray}
D& =& x_1x_3(1-r_2^{2})-r^{2}_{3}x^{2}_{3}\;,  \\
D_1& =& x_1x_3 + x_2x_3 - x_3+(-x_2^2 - x_1x_2 - x_3x_2 + 2x_2 + x_1
- x_1x_3 + x_3 - 1)r_2^2 \nonumber\\&&
  + r_3^2(-x_3^2 - x_2x_3 + x_3)+\frac{1}{4}r_2^{2} \;,\\
D_2& =&  x_1x_3 - x_2x_3 +(-x_2^2 + x1x_2 + x3x_2 - x_1x3)r_2^2 +
  r_3^2(x_2x_3 - x_3^2)+\frac{1}{4}r_2^{2}
\;.
\end{eqnarray}
In the calculation of hard function,  we  reserve the power terms of
$r_2$ up  to $\mathcal{O}(r^{4}_{2})$, the power terms of $r_3$ up
to $\mathcal{O}(r^{2}_{3})$, as $r_2^{4}$ and $r_3^{2}$ are taken as
zero, the hard function in this paper is same as  the hard function
in ref.\cite{cljpsi}.

 The hard scales $t$ are chosen as
\begin{eqnarray}
t^{(j)}={\rm max}(\sqrt{D}m_B,\sqrt{|D_j|}m_B,1/b_1)\;.
\end{eqnarray}

Similarly, we can get  the amplitude $A$ for $B^0\to J/\Psi
\eta^{\prime}$.  From the the amplitudes  for $B^0\to J/\Psi \eta$
and for $B^0\to J/\Psi \eta^{\prime}$, we can  derive the relation
between the amplitude $A$ for $B^0\to J/\Psi \eta^{\prime}$ and that
for $B^0\to J/\Psi \eta$ with the assumption that the decay constant
and the distribution amplitude  of the $\bar{d}d$ component for
$\eta$ is the same as that for
$\eta^{\prime}$\cite{feldmann1,feldmann,ball},
\begin{equation}\label{retapwitheta}
    A(B^0\to J/\Psi~\eta^{\prime})= \frac{\frac{1}{\sqrt{6}}\sin{\theta_P}+\frac{1}{\sqrt{3}}\cos{\theta_P}}{\frac{1}{\sqrt{6}}\cos{\theta_P}-\frac{1}{\sqrt{3}}\sin{\theta_P}}A(B^0\to J/\Psi \eta)
\end{equation}

\subsection{Numerical Analysis}
From the Eq.~(\ref{gamma}) and Eq.~(\ref{ampeta}), we can derive the
relation of the  branching ratio of $B^0 \to J/\Psi~\eta^{(\prime)}$
with $\eta-\eta^{\prime}$ mixing angle $\theta_{p}$,

\begin{eqnarray}\label{brbtoetatheta}
  Br(B^0\to J/\Psi~\eta) &=& \frac{1}{32\pi M_B \Gamma_{B^0}}G_F^2(1-r_2^2+\frac{1}{2}r_{2}^4-r_{3(\eta)}^2)\nonumber\\
                         &&({\frac{1}{\sqrt{6}}\cos{\theta_P}-\frac{1}{\sqrt{3}}\sin{\theta_P}})^2
  |(A_{FA}+ A_{VERT}+ A_{HS})|^2
\end{eqnarray}

\begin{eqnarray}\label{brbtoetathetap}
  Br(B^0\to J/\Psi~\eta^{\prime}) &=& \frac{1}{32\pi M_B \Gamma_{B^0}}G_F^2(1-r_2^2+\frac{1}{2}r_{2}^4-r_{3(\eta^{\prime})}^2)\nonumber\\
                                  &&( \frac{1}{\sqrt{6}}\sin{\theta_P}+\frac{1}{\sqrt{3}}\cos{\theta_P})^2|(A_{FA}+ A_{VERT}+ A_{HS})|^2
\end{eqnarray}
where $r_{3(\eta^{\prime})}=m_{\eta^{\prime}}/m_{B}$,
$r_{3(\eta)}=m_{\eta}/m_{B}$, $\Gamma_{B^0}$ is the total decay
width of $B^0$ meson.

\begin{figure}[htb]
\begin{center}
\psfig{file=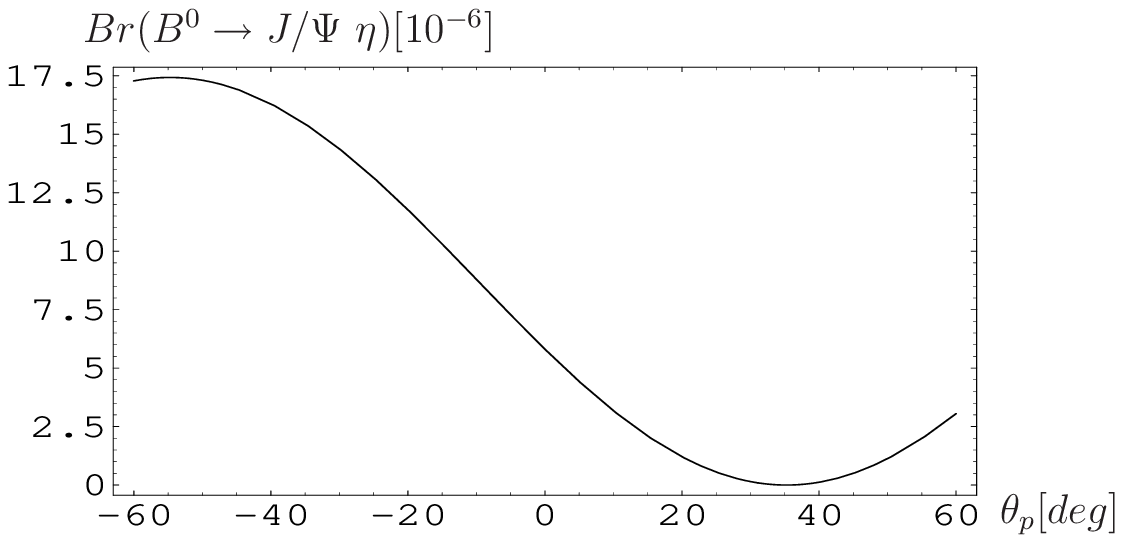,height=5cm,width=10cm}

 \end{center}
 \caption{The variation of
 the branching ratios of $B^0\to J/\Psi~\eta$ with $\eta-\eta^{\prime}$ mixing angle $\theta_{p}$
}\label{brbtoeta}
\end{figure}

\begin{figure}
\begin{center}
\psfig{file=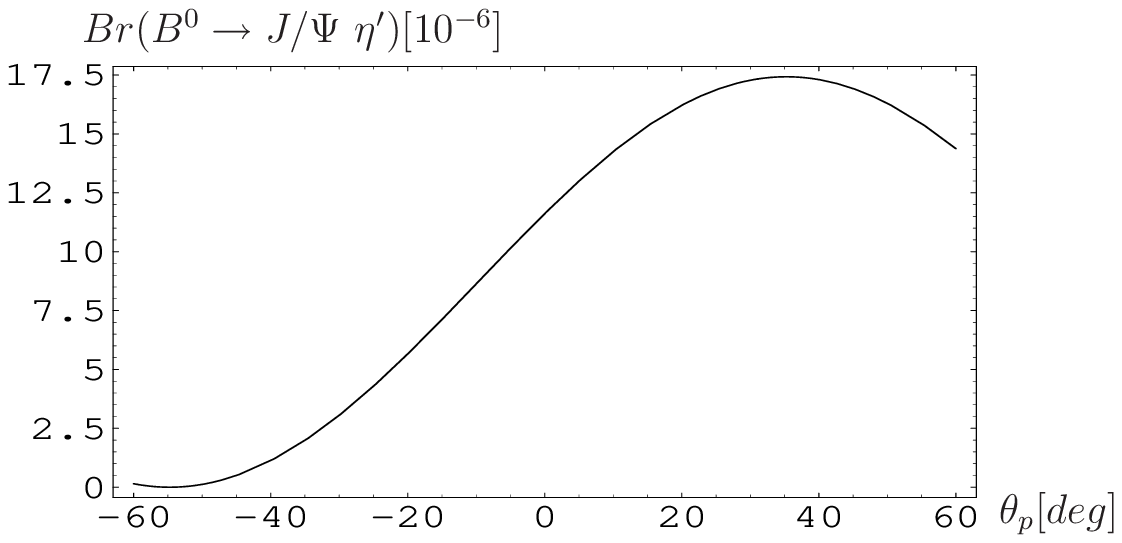,height=5cm,width=10cm}

 \end{center}
 \caption{The variation of
 the branching ratios of $B^0\to J/\Psi~\eta^{\prime}$ with $\eta-\eta^{\prime}$ mixing angle $\theta_{p}$
}\label{brbtoetap}
\end{figure}

The Fig.~(\ref{brbtoeta}) and  Fig.~(\ref{brbtoetap}) show the
relation of
 the branching ratios of $B^0\to J/\Psi~\eta^{(\prime)}$ with $\eta-\eta^{\prime}$
 mixing angle $\theta_{p}$ .

 According to Eq.~(\ref{brbtoetatheta}) and Eq.~(\ref{brbtoetathetap})
 we can determine the $\eta-\eta^{\prime}$ mixing
angle with the help of  the experimental data of the branching ratio
of $B^0 \to J/\Psi~\eta^{(\prime)}$. Because the the branching ratio
of $B^0 \to J/\Psi~\eta^{\prime}$ has not been measured, we try to
determine $\eta-\eta^{\prime}$ mixing angle $\theta_{p}$ according
to Eq.~(\ref{brbtoetatheta}) and  compare with the results of
others.   The range of $\theta_{p}$ is usually taken to be
$-20^{\circ}\leq\theta_{p}\leq -10^{\circ}$ in the literature. We
here choose the the range of $\theta_{p}$ as
$-60^{\circ}\leq\theta_{p}\leq 60^{\circ}$. Then we can get the
range of $\eta-\eta^{\prime}$ mixing angle $\theta_{p}$ from
Eq.~(\ref{brbtoetatheta}) and the recent experimental
data\cite{bellejpsi}
\begin{equation}\label{belldata}
    Br^{exp}(B ^0\to J/\Psi~\eta)=(9.6\pm1.7\pm0.7)\times 10^{-6},
\end{equation}


\begin{figure}[htb]
\begin{center}
\psfig{file=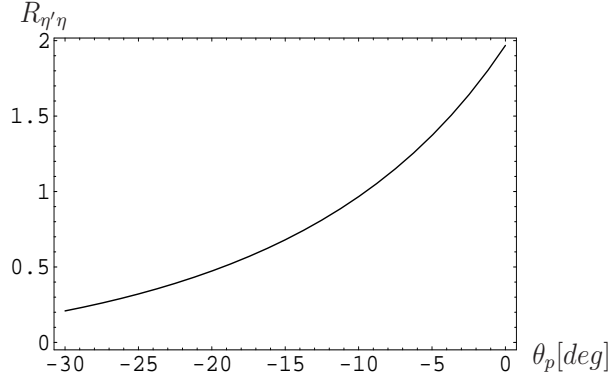,height=5cm,width=8cm}
\end{center}
 \caption{The variation of the
ratio for the branching ratios of $B^0\to J/\Psi~\eta^{\prime}$ and
$B^0\to J/\Psi~\eta$ with $\eta-\eta^{\prime}$ mixing angle
$\theta_{p}$ }\label{vretapeta}
\end{figure}
 Taking  the central value of experimental data in \cite{bellejpsi}, we
can get the $\eta-\eta^{\prime}$ mixing angle $\theta_{p}$
\begin{equation}\label{thetap}
\theta_{p}=-13.1^{\circ}
\end{equation}

Compared with the range of $\eta-\eta^{\prime}$ mixing angle
$-20^{\circ}\leq\theta_{p}\leq -10^{\circ}$ \cite{ekou01}, and the
recent result $\theta_{p}=-13.5^{\circ}$ or
$\theta_{p}=-17^{\circ}\pm1^{\circ}$ \cite{{kloenew,escribano,ht}},
our result is in agreement with theirs, but the $\theta_{p}$
determined in this method has some uncertainties induced by the
experimental data, CKM matrix element, $\eta^{(\prime)}$wave
functions, decay constants of $\eta^{(\prime)}$, form factor of $B
\to \eta$ transition, etc. So it is not a clean method.

 On the other hand, if we take the
 $\eta-\eta^{\prime}$ mixing angle as an input, say, $-20^{\circ}\leq\theta_{p}\leq
-10^{\circ}$, we can predict the branching ratio of $B ^0\to
J/\Psi~\eta$,
\begin{equation}\label{brbtojpsieta}
   8.68 \times 10^{-6}\leq Br(B ^0\to J/\Psi~\eta)\leq 11.6\times 10^{-6}
\end{equation}
Comparing our  result  in Eq.~(\ref{brbtojpsieta}) with experimental
data in Eq.~(\ref{belldata}) and that in Ref.\cite{xiaoeta} , \beq
Br(\ B^0 \to J/\Psi \eta) &=& \left [1.96^{+0.71}_{-0.50}(\omega_b)
 ^{+9.65}_{-0.39}(a_t)^{+0.32}_{+0.13}(a_2)^{+0.14}_{-0.13}( f_{J/\Psi})\right ] \times 10^{-6}, \label{eq:brje1}
  \label{eq:brjp1}
 \eeq
and other works in B decays into charmonia, we can conclude  that
pQCD approach can not apply to calculate the factorizable diagram
  in B decays into charmonia, because the characteristic hard scale is
  not big enough.

Now we discuss the other method for determining  the mixing angle
$\theta_{p}$. Because the branching ratio of $B^0\to
J/\Psi~\eta^{\prime}$ has not been measured, we need to calculate
the branching ratio of $B^0\to J/\Psi~\eta^{\prime}$
 . Taking the  $\eta-\eta^{\prime}$ mixing angle
 as $\theta_{p}=-13^{\circ}$, the branching ratio of $B^0\to
 J/\Psi~\eta^{\prime}$ can be  gotten from Eq.~(\ref{brbtoetathetap}),
 \begin{equation}\label{brbtoetap2}
    Br(B ^0\to J/\Psi~\eta^{\prime})=7.59\times 10^{-6}
 \end{equation}

 From
Eq.~(\ref{retapwitheta}), we can get the relation of the ratio of
the branching ratios of $B^0\to J/\Psi~\eta$ and $B^0\to
J/\Psi~\eta^{\prime}$ with the $\eta-\eta^{\prime}$ mixing angle
$\theta_{p}$,

\begin{equation}\label{rbretapwitheta}
    R_{\eta^{\prime}\eta}=\frac{Br(B^0\to J/\Psi~\eta^{\prime})}{Br(B^+\to J/\Psi \eta)}=
    \frac{(\frac{1}{\sqrt{6}}\sin{\theta_P}+\frac{1}{\sqrt{3}}\cos{\theta_P})^2}{(\frac{1}{\sqrt{6}}\cos{\theta_P}-\frac{1}{\sqrt{3}}\sin{\theta_P})^2}
    \frac{(1-r_2^2+\frac{1}{2}r_{2}^4-r_{3(\eta^{\prime})}^2)}{(1-r_2^2+\frac{1}{2}r_{2}^4-r_{3(\eta)}^2)}\;,
\end{equation}

 From the Eq.~(\ref{rbretapwitheta}), the mixing angle
$\theta_{p}$ can be extracted from the ratio of the branching ratios
of $B^0\to J/\Psi~\eta^{(\prime)}$. Because the uncertainty of the
masses of $\eta$ and $\eta^{\prime}$ is very small, the uncertainty
of $\theta_{p}$ determined in this method is mainly from the
uncertainty of the measured  ratio of the branching ratios of
$B^0\to J/\Psi~\eta^{(\prime)}$. It is obvious that the second
method can reduce the uncertainty quite a bit. In
Fig.(\ref{vretapeta}) we show the variation of the ratio of the
branching ratios of $B^0\to J/\Psi~\eta^{(\prime)}$ with
$\theta_{p}$. If the ratio of the branching ratios of $B^0\to
J/\Psi~\eta^{(\prime)}$ with $\theta_{p}$ were measured, we could
determine the mixing angle $\theta_{p}$ fairly well. We hope that
the future experiments would do it!

\section{Summary and Discussion}
In this paper, we derive the decay amplitude of $B^0\to
J/\Psi~\eta^{(\prime)}$ and the relation of the branching ratios of
$B^0\to J/\Psi~\eta^{(\prime)}$. We computed the branching ratio of
$B^0\to J/\Psi~\eta$ which is in agreement with recent experimental
data. We also predict the branching ratio of $B^0\to
J/\Psi~\eta^{(\prime)}$ to be $7.59\times 10^{-6}$. From the result
of the branching ratios of $B^0\to J/\Psi~\eta^{(\prime)}$ shown in
Table \ref{nfvchs1},\ref{nfvchs2}, we can find the nonfactorizable
contribution in $B^0\to J/\Psi~\eta^{(\prime)}$ is comparable to the
factorizable part, which is similar with in other B decay modes into
charmonia. Comparing many calculations to B decays into charmonia,
we conclude that  the pQCD approach can not be used to calculate the
factorizable contribution in B decays into charmonia. We suggest two
methods to determine the mixing angle $\theta_{p}$ of $\eta$ and
$\eta^{\prime}$. For the first method we get $\eta-\eta^{\prime}$
mixing angle $\theta_{p}$ to be about $-13.1^{\circ}$ which is in
consistency with others. The second method for determining the the
mixing angle $\theta_{p}$ can reduce the uncertainties  quite a bit,
but needs the experimental data of the branching ratio of $B^0\to
J/\Psi~\eta^{\prime}$ as an input. We hope that the future
experiment would measure it.

For comparison of the different contributions to the braching ratios
from naive factorization, vertex correction, and hard spectator
scattering, we present Table  \ref{nfvchs1},\ref{nfvchs2}. Form
Table  \ref{nfvchs1},\ref{nfvchs2} we can see that the vertex
correction and the spectator scattering are very important. The
naive factorization alone can not fit the data.

\begin{center}
\begin{table}[h]
\tabcolsep0.2in
\begin{tabular}{|c|c|c|c|}
\hline\hline
  quantity & NF& NF+VC & NF+VC+HS \\\hline
 Br($B\to J/\Psi~\eta$)$(10^{-6})$ & 1.615 & 2.719& 9.79\\\hline
Br($B\to J/\Psi~\eta^{\prime})$$(10^{-6})$ & 1.256 & 2.129 &
7.59\\\hline
  \hline
\end{tabular}
\caption{The branching ratios in units of $10^{-6}$ for $B\to
J/\Psi~\eta^{(\prime)}$, NF in Column II means the results in naive
factorization assumption, +VC in Column III, +HS in Column IV mean
the inclusions of with vertex correction in QCDF, hard spectator
contribution in pQCD, respectively.} \label{nfvchs1} \vskip 0.5cm
\begin{tabular}{|c|c|c|c|}
\hline\hline
  mode &$ A_{NF}$$(10^{-3})$& $A_{VC}$$(10^{-3})$& $A_{HS}$$(10^{-3})$ \\\hline
 $B\to J/\Psi~\eta$ &  -1.897-0.0584I & 3.658+1.962I&  -2.277+2.916I\\\hline
   $B\to J/\Psi~\eta^{\prime}$ & -1.592-0.0386I &  3.252+1.744I & -2.024+2.681I\\\hline
  \hline
\end{tabular}
\caption{The decay amplitude  in units of $10^{-3}$ for $B\to
J/\Psi~\eta^{(\prime)}$. $ A_{NF}$ in Column II means the
contribution of the factorizable diagram in naive factorization
assumption, $A_{VC}$ in Column III, $A_{HS}$ in Column IV mean the
contributions of  the vertex correction in QCDF, of the hard
spectator diagrams in pQCD, respectively.} \label{nfvchs2}
\end{table}
\end{center}
\section{Appendix: Input Parameters And Wave Functions}
We use the following input parameters in the numerical calculations
\beq
 \Lambda_{\overline{\mathrm{MS}}}^{(f=4)} &=& 250 {\rm MeV}, \quad
 \quad f_B = 190 {\rm MeV}, \quad  M_B = 5.2792 {\rm GeV}, \non
 M_W &=& 80.41{\rm GeV},\quad
\tau_{B^0}=1.53\times 10^{-12}{\rm
 s},
 \label{para}
\eeq
   For the CKM matrix elements,  we adopt the wolfenstein
parametrization for the CKM matrix up to $\mathcal{O}$$(\lambda^
3)$,
\begin{equation}
V_{CKM}= \left(           \begin{array}{ccc}
          1-\frac{\lambda^2}{2} & \lambda & A \lambda^3 (\rho-i \eta)\\
          -\lambda & 1-\frac{\lambda^2}{2} & A \lambda^2 \\
          A \lambda^3 (1-\rho-i \eta )&-A \lambda^2 & 1
          \end{array} \right) , \label{vckm}
\end{equation}
with the parameters $\lambda=0.2272, A=0.818, \bar{\rho}=0.221$ and
$\bar{\eta}=0.340$.

For the $B$ meson distribution amplitude, we adopt the
model\cite{luy01,kls01}

\beq \phi_B(x,b) &=& N_B x^2(1-x)^2 \mathrm{exp} \left
 [ -\frac{M_B^2\ x^2}{2 \omega_{b}^2} -\frac{1}{2} (\omega_{b} b)^2\right],
 \label{phib}
\eeq

where $\omega_{b}$ is a free parameter and we take
$\omega_{b}=0.4\pm 0.05$ GeV in numerical calculations, and
$N_B=91.745$ is the normalization factor for $\omega_{b}=0.4$.

The $J/\psi$ meson asymptotic distribution amplitudes are given by
\cite{BC04}
\begin{eqnarray}
\Psi^L(x)&=&\Psi^T(x)=9.58\frac{f_{J/\psi}}{2\sqrt{2N_c}}x(1-x)
\left[\frac{x(1-x)}{1-2.8x(1-x)}\right]^{0.7}\;,\nonumber\\
\Psi^t(x)&=&10.94\frac{f_{J/\psi}}{2\sqrt{2N_c}}(1-2x)^2
\left[\frac{x(1-x)}{1-2.8x(1-x)}\right]^{0.7}\;,\nonumber\\
\Psi^V(x)&=&1.67\frac{f_{J/\psi}}{2\sqrt{2N_c}}\left[1+(2x-1)^2\right]
\left[\frac{x(1-x)}{1-2.8x(1-x)}\right]^{0.7}\;,\label{jda}
\end{eqnarray}
The wave function for $d\bar{d}$ components of $\eta^{(\prime)}$
meson are given as \beq \Phi_{\eta_{d\bar{d}}}(P,x,\zeta)\equiv
\frac{1}{\sqrt{2N_C}} \left [ \psl
\Phi_{\eta_{d\bar{d}}}(x)+m_0^{\eta_{d\bar{d}}}
\Phi_{\eta_{d\bar{d}}}^{P}(x)+\zeta m_0^{\eta_{d\bar{d}}} ( \vsl
\nsl - v\cdot n)\Phi_{\eta_{d\bar{d}}}^{t}(x) \right ], \eeq where
$P$ and $x$ are the momentum and the momentum fraction of
$\eta_{d\bar{d}}$, respectively.

For $\eta^{(\prime)}$ meson, distribute amplitude is taken as
\cite{ball}

\beq \Phi_{\eta_{d\bar{d}}}(x)&=&\frac{3}{\sqrt{2N_c}}f_xx(1-x)
\left\{ 1+a_2^{\eta_{d\bar{d}}}\frac{3}{2}\left [5(1-2x)^2-1 \right
]\right. \non &&\left. + a_4^{\eta_{d\bar{d}}}\frac{15}{8} \left
[21(1-2x)^4-14(1-2x)^2+1 \right ]\right \},  \non
\Phi^P_{\eta_{d\bar{d}}}(x)&=&\frac{1}{2\sqrt{2N_c}}f_x \left \{ 1+
\frac{1}{2}\left (30\eta_3-\frac{5}{2}\rho^2_{\eta_{d\bar{d}}}
\right ) \left [ 3(1-2x)^2-1 \right] \right.  \non && \left. +
\frac{1}{8}\left
(-3\eta_3\omega_3-\frac{27}{20}\rho^2_{\eta_{d\bar{d}}}-
\frac{81}{10}\rho^2_{\eta_{d\bar{d}}}a_2^{\eta_{d\bar{d}}} \right )
\left [ 35 (1-2x)^4-30(1-2x)^2+3 \right ] \right\} ,  \non
\Phi^t_{\eta_{d\bar{d}}}(x) &=&\frac{3}{\sqrt{2N_c}}f_x(1-2x) \non
 && \cdot \left [ \frac{1}{6}+(5\eta_3-\frac{1}{2}\eta_3\omega_3-
\frac{7}{20}\rho_{\eta_{d\bar{d}}}^2
-\frac{3}{5}\rho^2_{\eta_{d\bar{d}}}a_2^{\eta_{d\bar{d}}})(10x^2-10x+1)\right
],  \non \eeq with \beq a^{\eta_{d\bar{d}}}_2&=& 0.44, \quad
a^{\eta_{d\bar{d}}}_4=0.25,\quad
 f_x=0.130{\rm GeV},  \non
\rho_{\eta_{d\bar{d}}}&=&m_{\pi}/{m_0^{\eta_{d\bar{d}}}}, \quad
\eta_3=0.015,\quad \omega_3=-3.0.  \eeq


\end{document}